\documentclass[prb,aps,twocolumn,floats]{revtex4-1} 
\usepackage{ifpdf}
\newif\ifpdf\ifx\pdfoutput\undefined\pdffalse\else\pdfoutput=1\pdftrue\fi

\ifpdf        \usepackage[pdftex]{graphics}
        \usepackage{epstopdf}
  \else\usepackage{graphics}\fi

\usepackage{amsmath}
\usepackage{graphicx}
\usepackage{epsfig}
\usepackage{psfrag}
\usepackage{color}
\newcounter{mytempeqncnt}
\newlength{\upit}\upit=0.1truein

\newcommand{\ltappr}{{{\lower4pthbox{$<$} } \atop \widetilde{ \ \ \ }}} 
\newlength{\bxwidth}\bxwidth=1.5 truein 
 
\newcommand{\tr}{{\hbox{Tr}}}

\newcommand{\dg}{^{\dagger }} 
 
\newcommand{\kk}{\mathbf{k}}

\newcommand{\bS}{\mathbf{S}}

\newcommand{\eps}{\epsilon}

\newcommand{\hR}{\hat{g}}
\newcommand{\rarrow}{\rightarrow}
\newcommand{\tPsi}{\tilde{\psi}}
\newcommand{\tf}{\tilde{f}}
\newcommand{\zmatrix}[4]{\left[\!\begin{matrix}#1 & #2\cr #3&#4\end{matrix}\!\right]}
\newcommand{\vertvec}[2]{\left(\!\!\begin{array}{c}#1\cr#2\end{array}\!\!\right)} 
 
\newcommand{\vertvecFour}[4]{\left(\!\!\begin{array}{c}#1\cr#2\cr#3\cr#4\end{array}\!\!\right)}

\newlength{\figwidth} 
\newlength{\shift} 
\newcommand{\fg}[3] 
{
\begin{figure}[!tb] 
\vspace*{-0cm} 
\[ 
\includegraphics[width=\figwidth]{#1} 
\] 
\vskip -0.2cm 
\caption{\label{#2} 
\small#3 
}
\end{figure}} 
\newcommand{\fgs}[3] 
{
\begin{figure*}[ht] 
\vspace*{-0cm} 
\[ 
\includegraphics[width=16cm]{#1} 
\] 
\vskip -0.2cm 
\caption{\label{#2} 
\small#3 
}
\end{figure*}} 
\newcommand \bea {\begin{eqnarray} } 
\newcommand \eea {\end{eqnarray}} 
\newcommand \beq {\begin{equation} } 
\newcommand \eeq {\end{equation}} 
\newcommand{\bk}{{\bf{k}}} 
\newcommand{\bx}{{\bf{x}}}
\newcommand{\bR}{{\bf{R}}}

\newcommand{\ud}{\mathrm{d}}
\newcommand{\pd}{\partial}

\newcommand{\e}{{\mathrm{e}}}
\begin{document} 

\title{Composite pairing in a mixed valent two channel Anderson model}

\author{Rebecca Flint$^{1,3}$, Andriy H. Nevidomskyy$^{2,3}$ and Piers Coleman$^3$} 
\affiliation{$^1$Department of Physics, Massachusetts Institute of Technology, Cambridge MA 02139, USA}
\affiliation{$^2$Department of Physics and Astronomy, Rice University, Houston, TX 77005, USA}
\affiliation{$^3$Center for Materials Theory, Rutgers University, Piscataway, NJ 08854,
USA}

\begin{abstract} 
Using a two-channel Anderson model, we develop a theory of composite
pairing in the 115 family of heavy fermion superconductors that
incorporates the effects of $f$-electron valence fluctuations.  Our
calculations introduce ``symplectic Hubbard operators'': an extension
of the slave boson Hubbard operators that preserves both spin rotation
and time-reversal symmetry in a large $N$ expansion, permitting a
unified treatment of anisotropic singlet pairing and valence
fluctuations.  We find that the development of composite pairing in
the presence of valence fluctuations manifests itself as a
phase-coherent mixing of the empty and doubly occupied configurations
of the mixed valent ion.  This effect redistributes the $f$-electron
charge within the unit cell. Our theory predicts a sharp
superconducting shift in the nuclear quadrupole resonance frequency
associated with this redistribution. We calculate the magnitude and
sign of the predicted shift expected in CeCoIn$_5$.
\end{abstract} 
 
\maketitle

\section{Introduction}

The 115 family of superconductors continue to attract attention for
the remarkable rise in the superconducting transition temperature from
$0.2$~K in the parent compound CeIn$_{3}$[\onlinecite{mathur98}] to $2.3$~K in
CeCoIn$_5$[\onlinecite{petrovic01a}] and finally to $4.9$~K and $18.5$~K in
the actinide analogues, NpPd$_5$Al$_2$[\onlinecite{aoki08}] and
PuCoGa$_5$[\onlinecite{sarrao02}], respectively.  This last rise has often been
attributed to the increasing importance of valence fluctuations, and
here we seek to make this connection explicit.  Although the Ce 115s
are local moment systems, neutron studies of crystal fields indicate
broad quasi-elastic line-widths comparable to the crystal field
splitting, indicating valence fluctuations\cite{severing10}.  The
highest $T_c$ 115s, the actinides, involve 5$f$ shell electrons, which
are less localized than their 4$f$ Ce counterparts, and thus are
expected to be even more mixed valent.  Recent $^{237}$Np M\" ossbauer
studies of NpAl$_{2}$Pd$_{5}$ suggest that the valence of the Np ions
actually changes as superconductivity develops\cite{gofryk09}.
However, despite different degrees of mixed valency, both CeCoIn$_5$
and NpPd$_5$Al$_2$ contain the same unusual transition from local
moment paramagnetism directly into the superconducting state, shown in
Figure \ref{chi}.

This direct transition suggests that the localized moments play a
direct role in the pairing, which previously led us to propose that
the 115 materials are \emph{composite pair
superconductors}~\cite{flint08,flint10}.  Composite pair
superconductivity is a local phenomenon involving the condensation of
bound states between local moments and conduction electrons\cite{abrahams95}: it can be
thought of as an \emph{intra-atomic} version of a d-wave magnetic
pairing, between conduction electrons in orthogonal screening channels
rather than on neighboring sites,
\begin{equation}
\Delta_{SC}(j) = \langle \psi\dg_{1j} \vec{\sigma}(i\sigma_2) \psi\dg_{2j} \cdot \vec{S}_{fj}\rangle.
\end{equation}
Here $\psi_{1,2j}\dg$ create local Wannier states of conduction
electrons in two orthogonal symmetry channels at site $j$, and
$\vec{S}_{fj}$ is the local $f$-moment on the same site.  The
composite pair combines a triplet pair of conduction electrons with a
spin-flip of the local moment, such that the overall pair remains a
singlet.

While the close vicinity of the Ce 115s to antiferromagnetic order has
led to a consensus that pairing in these materials is driven by spin
fluctuations, the two superconducting domes in the Ce(Co,Rh,Ir)In$_5$
phase diagram\cite{pagliuso01} suggest that composite pairing may provide a second,
complementary mechanism in the Ce 115s\cite{flint10}.  The absence of
magnetism in the actinide 115 phase diagrams suggests that composite
pairing may play a more important role.  Composite pairs are predicted
to have unique electrostatic signature, resulting in a small
redistribution of the $f$-electron charge within the unit cell and an
associated change of the $f$-valence.  Understanding these charge
aspects of composite pairing is essential to disentangling the
relative importance of magnetic and composite pairing in these
compounds.

\fg{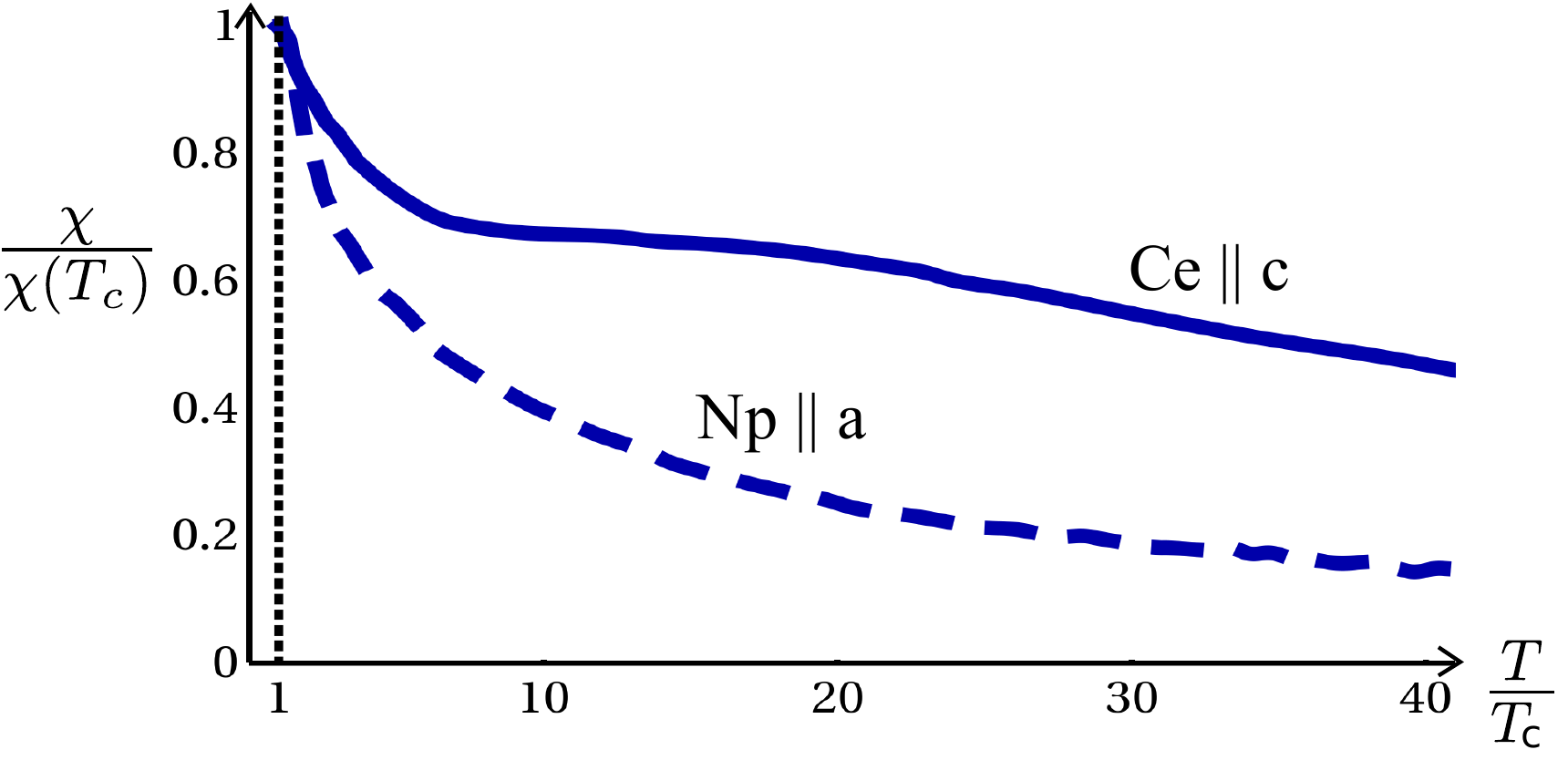}{chi}{Local moments are seen in the Curie-Weiss
susceptibilities: CeCoIn$_5$($T_c = 2.3K$)\cite{shishido02} and
NpPd$_5$Al$_2$($T_c = 4.9K)$\cite{aoki08} are reproduced and rescaled
by $\chi(T_c)$ to show their similarity (data below $T_c$ not shown).}

These observations motivate us to develop a theory of composite
pairing incorporating valence fluctuations.  Thus far, composite
pairing has been studied within a two-channel Kondo
model\cite{catk,flint08}, treating only the local spin degrees of
freedom.  In this paper, we study composite pairing within a
two-channel Anderson model, which permits us to include the charge
degrees of freedom and model the effects of valence fluctuations on
the superconductivity in the 5$f$ 115 materials.  We predict a sharp
shift in both the $f$-electron valence and quadrupole moment at the
superconducting transition temperature.  The quadrupole moment shift
should manifest as a shift either in the nuclear quadrupole resonance
(NQR) frequency, for Ce$M$In$_5$ or the M\"ossbauer quadrupole
splitting, for NpPd$_5$Al$_2$, and we make concrete predictions for
these shifts in Section ~\ref{Sec.NQR}.

The two-channel Anderson model is the natural extension of the two-channel Kondo model.
\begin{equation}
H = H_c + \sum_j H_a(j) + \sum_j H_V(j).
\end{equation}
Here the local atomic Hamiltonian at site $j$ is,
\begin{equation}
\label{atomic}
H_a(j) = E_0 X_{00}(j) + E_2 X_{22}(j) + \sum_\sigma \epsilon_f X_{\sigma \sigma}(j),
\end{equation}
where the $X$'s are the Hubbard operators\cite{hubbard64}: $X_{00} = |0\rangle \langle
0 |$, $X_{22} = |2\rangle \langle 2|$, and $X_{\sigma \sigma} =
|\sigma\rangle \langle \sigma|$.  The atomic states $|0\rangle$,
$|\sigma\rangle$ and $|2\rangle$ are shown in Figure \ref{2levels},
where we take the doubly occupied state to be a singlet containing
$f$-electrons in two orthogonal channels,
\begin{equation}
\vert 2 \rangle \equiv \vert \Gamma_{2} \otimes \Gamma_{1}\rangle_{s} =\frac{1}{\sqrt{2}} \sum_{\sigma= \pm 1
}{\rm sgn} (\sigma)f\dg_{\Gamma_{2}\sigma } f\dg_{\Gamma_{1} \ -\sigma }\vert 0\rangle .
\end{equation}
The orthogonality of these two channels is essential to the
superconductivity, and we take the Coulomb energy for two electrons in
the same channel to be infinite, making this model closer to two copies of the infinite-$U$ Anderson model than to the
finite-$U$ model\cite{bolech}.  These $f$-electrons hybridize with a bath of
conduction electrons, $H_c = \sum_{\bk,\sigma} \epsilon_\bk
c\dg_{\bk\sigma} c_{\bk\sigma}$ in two different channels
\begin{equation}  
\label{N2mix} H_V(j) = \sum_{j}V_{1}
\psi\dg_{1j\sigma} X_{0\sigma}(j) +\mathrm{h.c.} + V_{2} \psi\dg_{2j
\sigma} \tilde{\sigma}X_{-\sigma 2}(j) + \mathrm{h.c.}, 
\end{equation}  
where the $\psi_{\Gamma j \sigma}$ are Wannier states representing a
conduction electron in symmetry $\Gamma$ on site $j$.  $X_{0\sigma} =
|0\rangle \langle \sigma |$ and $X_{2\sigma} = |2\rangle \langle\sigma
|$ are the Hubbard operators between the singly occupied state and
empty or doubly occupied states, respectively (see Fig.~\ref{space}).  Here we have adopted
the notation of Ce atoms, whose ground state is a $4f^1$ doublet, but
this formalism also applies to Np $5f^3$ atoms, where both
$|0\rangle \equiv |5f^4\rangle$ and $|2\rangle \equiv |5f^2\rangle$
are $J=4$ crystal field singlets, while $|\sigma\rangle$ represents
one of the $J=9/2$ crystal field doublets of $5f^3$.

\fg{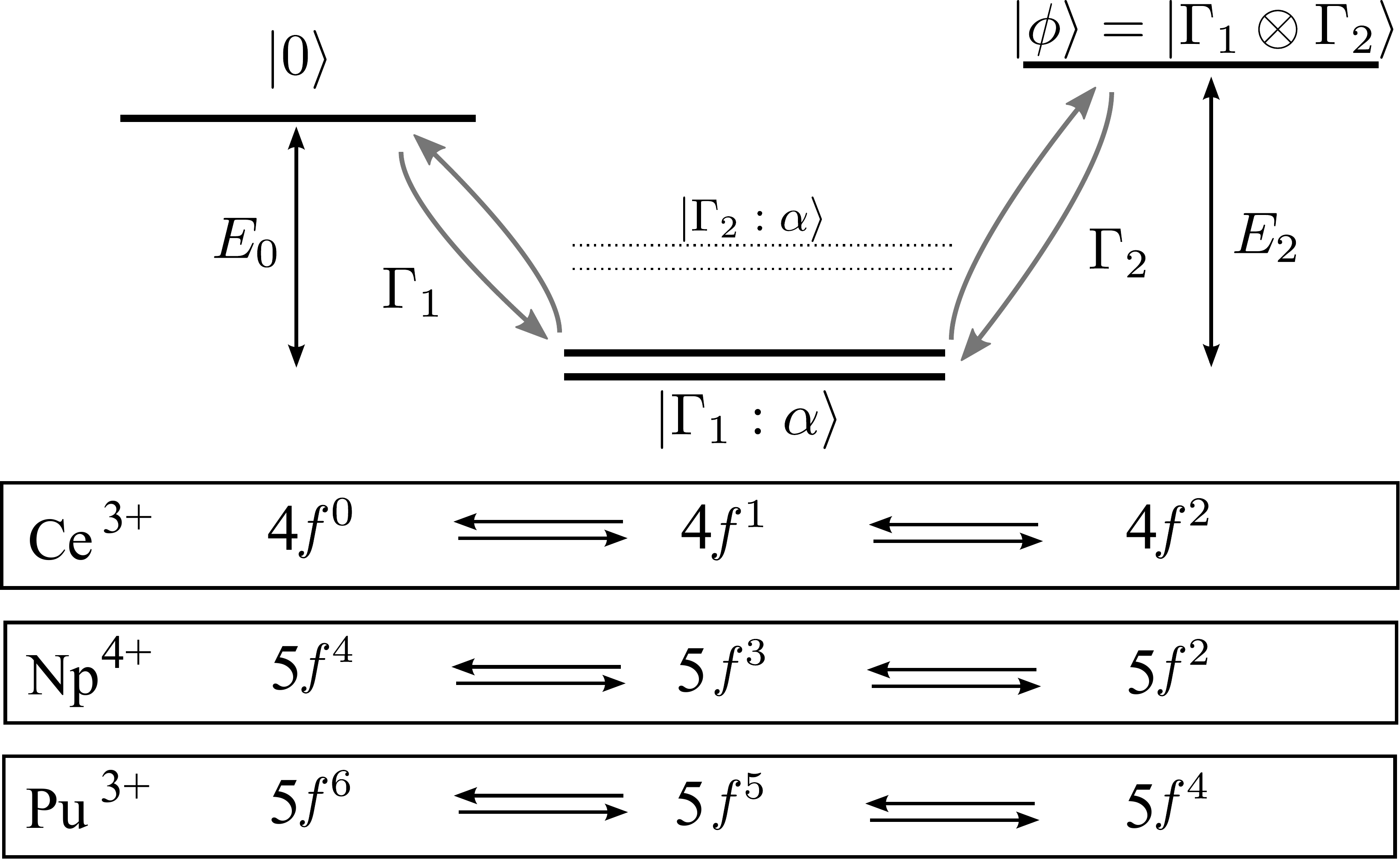}{2levels}{Virtual charge fluctuations of a
two channel Anderson impurity, where the addition and removal of an
$f$-electron occur in channels $\Gamma_{1}$ and $\Gamma_{2}$ of
different crystal field symmetry.  The ground state is a Kramer's
doublet, while the excited states $|0\rangle$, $|2\rangle$ are singlets.  The excited 
doublet $|\Gamma_2:\alpha\rangle$ represents a higher lying crystal-field level and 
is excluded from the Hilbert space of the problem.
The relevant charge states of
Ce$^{3+}$ (4$f^1$), Np$^{4+}$ (5$f^3$) and Pu$^{3+}$ (5$f^5$) are indicated. }

To develop a controlled treatment of superconductivity in the
two-channel Anderson model, we introduce a large-$N$ expansion.
Preserving the time-reversal symmetries of spins is essential to
capture singlet superconductivity in the entire family of large $N$
models, and large $N$ methods based on $SU(N)$ spins, like the usual
slave boson approach, will lose time-reversal symmetry for any $N >
2$.  Treating superconductivity in the Kondo model therefore requires
symplectic spins, the generators of the $SP(N)$ group\cite{flint08},
\begin{equation}\label{}
S_{\alpha \beta} = f\dg_\alpha f_\beta - \tilde{\alpha}\tilde{\beta}
f\dg_{-\beta}f_{-\alpha} 
.
\end{equation}
Here $N$ is an {\sl even} integer and the indices
\begin{equation}\label{}
\alpha\in \{\pm 1, \pm 2, \ldots \pm N/2 \}
\end{equation}
are integers running from $-N/2$ to $N/2$, excluding zero. We
employ the notation $
\tilde{\alpha} \equiv \mathrm{sgn}(\alpha)$.
These
spin operators invert under time-reversal $\bS \to \theta
\bS\theta^{-1}= -\bS$ and generate rotations that commute with the
time-reversal operator\cite{flint08}.

Physically, the spin fluctuations of a local moment are generated by
valence fluctuations. Theoretically, these valence fluctuations can be
described by Hubbard operators, and in a symplectic-$N$ generalization
of the Anderson model, anticommuting two such Hubbard operators must
generate a symplectic spin, satisfying the relations:
\begin{eqnarray}
\label{algebra}
\left\{X_{0\alpha},X_{\beta 0}\right\} &=& X_{\alpha\beta} + X_{00} \delta_{\alpha\beta} \\
&=& S_{\alpha \beta} + \left(X_{00}+\frac{X_{\gamma
\gamma}}{N}\right)\delta_{\alpha,\beta},\nonumber
\end{eqnarray}
where the last equality follows from the traceless definition of the
symplectic spin operator, $S_{\alpha \beta} = X_{\alpha \beta} -
\frac{X_{\gamma \gamma}}{N}\delta_{\alpha\beta}$.  In Section
\ref{Sec.Hubbard}, we show that a proper symplectic representation requires the
introduction of \emph{two} slave bosons to treat the Hubbard operators
in a single channel:
\begin{eqnarray}  X_{0\alpha} & = & b\dg f_\alpha + a\dg
\tilde{\alpha} f_{-\alpha}\dg\cr X_{00} & = & b\dg b + a\dg a.
\end{eqnarray}  
As we shall demonstrate, these symplectic Hubbard operators maintain
the invariance of the Anderson Hamiltonian with respect to the $SU(2)$
particle-hole transformation, $f_\sigma \to \cos\theta\, f_\sigma +
\mathrm{sign}(\sigma)\sin\theta\,f\dg_{-\sigma}$\cite{affleck88}.
This extension of the slave boson representation with local 
$SU (2)$ gauge symmetry
was originally introduced by Wen and Lee as a mean-field treatment of
the $t-J$ model\cite{wenlee96}. Here we show that these Hubbard operators
maintain the $SU(2)$ gauge symmetry for arbitrary $SP(N)$ versions of the
infinite $U$ Anderson and $t-J$ models.  This $SU(2)$ gauge symmetry is 
essential for eliminating the false appearance of s-wave
superconductivity in the finite-$U$ Anderson model, where the two
channels are identical.

What do we learn from this Anderson model picture of composite
pairing?  A key result is that the amount of composite pairing can be
written in a simple gauge-invariant form as \beq \Delta_\text{SC} =
\langle X_{02} \rangle \propto \langle \psi\dg_1 \vec{\sigma} \left(
i\sigma_2\right) \psi\dg_2 \cdot \vec{S}_f\rangle, \eeq where the
Hubbard operator $X_{02}=|0\rangle\langle 2|$ mixes the empty and
doubly occupied states.  This mixing resembles an intra-atomic version
of the negative-$U$ pairing state, but double occupancy costs a large,
positive energy.  The physical consequence of this mixing is a
redistribution of the charge in the $f$-electron orbitals, as the
three charge states: empty, singly and doubly occupied, all have
different charge distributions.  Such a rearrangement not only changes
the total charge in the $f$-shell, a monopole effect, but also will
result in a quadrupole moment associated with the superconductivity.
This quadrupole moment will lead to modified electric field gradients
that should be detectable as a sharp shift in the nuclear quadrupolar
resonance (NQR) frequency at the neighboring nuclei, or as a sharp
shift in the M\"{o}ssbauer quadrupolar splitting at the $f$-electron
nuclei.  

This paper is organized as follows. First we introduce the symplectic
Hubbard operators in section \ref{Sec.Hubbard} and demonstrate that
the symplectic-$N$ and $SU(N)$ large-$N$ limits are identical for a
single channel.  We then generalize this formalism to the case of two
channels in Section~\ref{Sec.2ch} and show how composite pairing
naturally appears as a mixing of the empty and doubly occupied
states. The mean-field solution is presented in Section~\ref{Sec.MF},
in which we demonstrate how the superconducting transition temperature
increases with increasing mixed valence.  In Section~\ref{Sec.NQR}, we
calculate the charge distribution of the $f$-orbitals in the state
with composite pairing and make a concrete prediction for a shift in
the NQR frequency at $T_c$ in CeCoIn$_5$.  Finally, section \ref{Sec.Concl}
discusses the implications for the finite-$U$ Anderson model and
examines the broader implications of our results.

\section{Symplectic Hubbard operators}
\label{Sec.Hubbard}

Composite pairing was originally discussed in the two channel Kondo model\cite{catk}, where it is found in
the symplectic-$N$ limit\cite{flint08}, which maintains the time-inversion
properties of spins in the large $N$ limit by using symplectic spins.
Correctly including time-reversal symmetry allows the formation of
Cooper pairs, and thus superconductivity.  In order to treat composite
pairing within a large $N$ Anderson model treatment, we would like to
develop a set of Hubbard operators that maintain this time-reversal
property in the large $N$ limit.  Hubbard operators, like $X_{ab} =
|a\rangle \langle b|$, are projectors between the local states,
$|0,\sigma\rangle$ of the infinite $U$ Anderson model, describing both
charge, $X_{0 \sigma}$ and spin, $X_{\sigma \sigma'}$ fluctuations.

Starting from a spin state $|\sigma\rangle$, hopping an electron off
and back onto the site generates a spin flip: this condition
defines the Hubbard operators within a single channel, where the
Hubbard operators must satisfy a graded Lie algebra in which the
projected hopping operators anti-commute, satisfying the algebra
\begin{equation}\label{opalg}
\left\{X_{\alpha 0}, X_{0\beta}
\right\} = 
X_{\alpha \beta }+X_{00}\delta_{\alpha \beta }
\end{equation}
Now the traceless part of $X_{\alpha \beta  }$ defines a spin
operator,
\[
S_{\alpha \beta } = X_{\alpha \beta } - \frac{\delta_{\alpha \beta }}{N}X_{\gamma\gamma}
\]
so that quite generally, Hubbard operators must satisfy an algebra of
the form 
\begin{equation}
\left\{X_{\alpha 0},X_{0\beta}
\right\} 
= S_{\alpha \beta} + \left(X_{00}+\frac{X_{\gamma \gamma}}{N}\right)\delta_{\alpha\beta}
\end{equation}
Thus the commutation algebra of the Hubbard operators generates
the spin operators of the local moments. 
In the 
traditional slave boson approach\cite{coleman83}, 
the Hubbard operators are written
$X_{0\alpha} = b\dg f_\alpha$.  
The spin operators generated by this procedure
are the generators of $SU (N)$, since 
\begin{equation}\label{}
\{f\dg_{\alpha }b, b\dg f_{\beta } \} = 
\overbrace {(f\dg_{\alpha }f_{\beta } - \frac{n_{f}}{N}\delta_{\alpha
\beta })}^{
S^{SU(N)}_{\alpha \beta}} + 
\left(b\dg b + \frac{n_{f}}{N} \right)\delta_{\alpha \beta}
\end{equation}
where 
$S^{SU (N)}_{\alpha \beta}$ is  the well-known form of
$SU(N)$ spins.  
It is thus no wonder that this approach cannot treat
superconductivity in the large $N$ limit, for there are no
spin singlet Cooper pairs for $SU(N)$ with $N>2$.
We require instead that the
spin fluctuations generated by Hubbard operators are symplectic spin operators
\begin{equation}\label{}
S_{\alpha \beta } = f\dg_{\alpha }f_{\beta }- {\rm sgn} (\alpha \beta
) f\dg_{-\beta}f_{-\alpha}
\end{equation}

We now show that the Hubbard algebra(\ref{algebra})
can be satisfied with symplectic spins, 
 using the introduction of
\emph{two} slave bosons,
\begin{eqnarray} 
X_{\alpha 0} & =  & f\dg _{\alpha }b + \tilde{\alpha} f_{-\alpha}a,\cr
X_{0\beta} & =  & b\dg f_\beta + a\dg \tilde{\beta} f_{-\beta}\dg .
\end{eqnarray}
These operators satisfy the Hubbard operator algebra (\ref{opalg}),
but now
\begin{equation}\label{}
X_{\alpha \beta }= f\dg_{\alpha }f_{\beta }+ {\tilde{\alpha}
\tilde{\beta }}f_{-\alpha }f\dg_{-\beta } = S_{\alpha \beta }+ \delta_{\alpha \beta }
\end{equation}
describes a symplectic spin operator while
\begin{eqnarray}\label{l}
X_{00}  =   b\dg b + a\dg a 
\end{eqnarray}
is the representation of the empty state operator.

At first sight, the expedience of this new representation might be
questioned: why exchange the simplicity of the original 
Hubbard operators, $X_{ab}
=|a\rangle \langle b|$ for a profusion of slave boson fields?
However, while the original Hubbard operators may appear to be simple, they are
singularly awkward to treat in many-body theory
\cite{keiter71}, due to their noncanonical anti-commutation
algebra, (\ref{algebra}).  The slave boson representation
allows us to represent these operators in terms of
canonical bosons and fermions.  By doubling the number of slave bosons
the symplectic character of the spins is preserved at all even values
of $N$, and we shall see that
this process encodes the hard-to-enforce Gutzwiller projection as a 
mathematically tractable $SU(2)$ gauge symmetry.

Physical spins are \emph{neutral}, and thus possess a continuous
particle-hole symmetry.  This property is maintained by symplectic
spin operators, which can be seen most naturally by introducing the
generalized pair creation operators,
\begin{equation}
\Psi\dg  = \frac{1}{2}\sum_\alpha \tilde{\alpha} f\dg_\alpha f\dg_{-\alpha} = \sum_{\alpha >0} f\dg_\alpha f\dg_{-\alpha},
\end{equation}
which allow us to construct the isospin vector, $\vec{\Psi} = (\Psi_1, \Psi_2, \Psi_3)$
, where
\begin{eqnarray} 
\label{isospin}
\Psi_1 & = & \left( \Psi\dg + \Psi\right), \; \; 
\Psi_2 = -i \left(\Psi\dg - \Psi\right) \cr
\Psi_3 & = & \sum_{\alpha >0}(f\dg_\alpha f_\alpha - f_{-\alpha}f\dg_{-\alpha}) = n_s -N/2,
\end{eqnarray} 
and $n_{s} \equiv \sum_\alpha \langle\alpha|f\dg_\alpha
f_\alpha|\alpha\rangle$ is the fermion number in the
singly-occupied states, i.e. $n_s$ plays the role of the number of
spins in the above constraint $\Psi_3$.  The isospin vector 
commutes with symplectic spins, $\left[ \vec{\Psi}, S_{\alpha
\beta}\right] = 0$, showing that the symplectic spins possess
an $SU(2)$ gauge symmetry: a continuous particle-hole symmetry that
allows us to redefine the spinon, $f_\alpha \rarrow u f_\alpha + v
\tilde{\alpha} f\dg_{-\alpha}$.  This symmetry is reflected in the
requirement of two types of bosons, as the empty state does not
distinguish between zero and two fermions, and thus requires two
bosons to keep track of the two ways of representing the empty state,
$b\dg|\Omega\rangle$ and $a\dg f\dg_\uparrow
f\dg_\downarrow|\Omega\rangle =a\dg \Psi\dg|\Omega\rangle$, where
$|\Omega\rangle$ is the slave-boson vacuum.  Of course, there is only one physical
empty state, as becomes clear when we restrict these Hubbard operators
to the physical subspace.  In order to faithfully represent the
symplectic spins, the sum of the spin and charge fluctuations must be
fixed, $\vec{S}^2 + \vec{\Psi}^2 = \frac{N}{2}(\frac{N}{2} +
2)$[\onlinecite{flint08}].  While this constraint is
enforced by setting $\vec{\Psi} =0$ in the pure spin model, here we must equate our two types
of charge fluctuations, by setting
\begin{eqnarray} 
Q_3 & = & \sum_{\alpha} f\dg_\alpha f_\alpha - N/2 + b\dg b - a\dg a = 0 \cr
Q_+ & = & \sum_{\alpha > 0} f_\alpha\dg f_{-\alpha}\dg + b\dg a = 0 \cr
Q_- & = & \sum_{\alpha > 0} f_{-\alpha} f_{\alpha} + a\dg b  = 0.
\end{eqnarray} 

The three operators $(Q_{\pm},Q_{3})$ commute with the Hamiltonian,
which imposes the constraints associated with the 
physical Hilbert space. The constraint reflects the neutrality of the spins under charge conjugation: $Q_3$ conserves total electromagnetic charge, and prevents double occupancy, while $Q_\pm$ kills any states with s-wave pairs on-site.
From the form of $Q_{3}$, we see that 
$b$ and $a$ have opposite gauge charges. The gauge invariant states satisfying the constraint (for $N = 2$) are,
\begin{eqnarray} 
|\alpha\rangle & = & f\dg_\alpha |\Omega\rangle\cr
|0\rangle & = &\left(b\dg + a\dg \Psi\dg\right)|\Omega\rangle.
\end{eqnarray} 

The symplectic Hubbard operators can be written more compactly by using Nambu notation,
\begin{eqnarray} 
X_{0\alpha}  = B\dg \tilde{f}_\alpha, \; X_{00} = B\dg B 
\end{eqnarray}
where 
\begin{eqnarray} 
B = \left(\!\begin{array}{c}
b \cr
a
\end{array}\!
\right),
\; \tilde{f}_\alpha =  \left(\!\begin{array}{c}f_\alpha\\
\tilde{\alpha} f_{-\alpha}\dg\end{array}\!\right), 
\end{eqnarray} 
The constraint becomes $\vec{Q} = B\dg \vec{\tau} B + \sum_{\alpha>0}\tilde{f}_\alpha\dg \vec{\tau} \tilde{f}_\alpha = 0$.
For $N = 2$, these Hubbard operators are the $SU(2)$ slave bosons
introduced by Wen and Lee in the context of the $t-J$
model\cite{wenlee96}.  Here, it becomes clear that the $SU(2)$
structure is a consequence of symplectic symmetry, present in both the
symplectic-$N$ Kondo and Anderson models, not just for $N=2$, but 
for all $N$.  This symmetry
can be physically interpreted as the result of valence fluctuations in
the presence of a particle-hole symmetric spin.  

The Hamiltonian of the one channel, single impurity infinite-$U$ Anderson model can now be expressed via symplectic slave bosons as follows:
\begin{eqnarray} 
H & = & \sum_{\bk,\alpha} \epsilon_\bk c\dg_{\bk\alpha } c_{\bk\alpha}
+ E_0 (b\dg b + a\dg a)\cr & +& \sum_{\alpha}V \psi\dg_{\alpha}(b\dg f_{\alpha} + a\dg \tilde{\alpha}
f_{-\alpha}\dg)+\mathrm{h.c.} \label{H1}
\end{eqnarray} 
The $SU$(2) gauge symmetry of this model becomes particularly evident
in the field-theoretical formulation, where the partition function is
written as a path integral: 
\beq Z = \int
\mathcal{D}[\tPsi\dg,\tPsi,\tf\dg,\tf,B\dg,B] e^{-S}, \eeq where
the Euclidean action $S$ is written in Nambu notation as follows
\setcounter{mytempeqncnt}{\value{equation}}
\begin{widetext}
\setcounter{equation}{0}
\begin{equation}
S
=\! \int\limits_0^\beta\! \ud\tau\!
\sum_{\kk,{\alpha >0}} \tPsi_{\kk{\alpha}}\dg (\pd_\tau+\eps_k
\tau_3)\tPsi_{\kk{\alpha}} + \! \sum_{j}\! \left\{
\frac{1}{2}
\tr\left[ \tau_3 \mathcal{A}\dg_{j}(\pd_\tau\! + E_0\! +
\vec{\lambda}_j\cdot\vec{\tau})\mathcal{A}_{j}
\right] 
+
\sum_{\alpha >0}\left[
\tf\dg_{j{\alpha}}(\pd_\tau +
\vec{\lambda}_j\cdot\vec{\tau})\tf_{j{\alpha}} \nonumber 
+ \left(V
\tPsi\dg_{j\alpha}\mathcal{A}_j\dg\tf_{j\alpha} + \text{h.c.}\right)
\right]
\!
\right\}\!, \tag{27}
\label{action1}
\end{equation}
\end{widetext}
\addtocounter{mytempeqncnt}{2}
\setcounter{equation}{\value{mytempeqncnt}}
where
$\tPsi\dg_\alpha = (\psi\dg_\alpha, \tilde{\alpha} \psi_{-\alpha})$,
$\beta=1/T$ is the inverse temperature and the $SU$(2) matrix
$\mathcal{A}_j$ of slave bosons is defined as 
\begin{eqnarray}
\mathcal{A}_j = \zmatrix{b_j}{a_j\dg}{a_j}{-b_j\dg }.
\end{eqnarray}
The first three terms in Eq.~(\ref{action1}) describe the dynamics of conduction electrons $\tPsi$, $f$-electrons $\tf$ and the slave bosons $B$ respectively, with the constraint $\vec{Q}=0$ imposed by introducing the vector of Lagrange multipliers $\vec{\lambda}$. The last term describes the coupling between the f-atom and conduction electrons.

The action (\ref{action1}) is manifestly invariant under the $SU(2)$ gauge transformation via unitary matrix $\hR_j$:
\bea
\tf_j \rarrow \hR_j\tf_j,&& \quad \tPsi_j \rarrow \hR_j \tPsi_j, \quad \mathcal{A}_j \rarrow \hR_j \mathcal{A}_j \\
&&\vec{\lambda_j}\cdot\vec{\tau} \rarrow \vec{\lambda_j}\cdot\vec{\tau} - \hR_j\dg \pd_\tau \hR_j \nonumber
\eea
At first sight, the $a$ hybridization term in the Hamiltonian,
\eqref{H1} appears to give rise to $\langle c\dg f\dg\rangle$ pairing
terms, however, the $SU(2)$ symmetry allows to redefine the fields,
making the substitution
\beq
B=  \binom{b}{a} \rarrow \hR B = \binom{b'}{0}.
\eeq
This transformation eliminates the composite $s$-wave pairing, $b\dg f_\alpha + a\dg
f\dg_{-\alpha} \tilde{\alpha} \rightarrow b^{'\dagger} f'_\alpha$,
recovering the usual $U(1)$ slave boson Hamiltonian.

One of the the most important physical consequences encoded in this
\emph{local} gauge symmetry is the complete suppression of 
any $s$-wave pairing in the single channel Anderson model. Indeed, the constraint 
operator $Q_{+} = \sum_{\alpha>0} f\dg_\alpha f\dg_{-\alpha}+ b\dg a$ may be directly interpreted
as the on-site $s$-wave pair creation operator plus a term $b\dg a$
that can be removed by the above gauge transformation. In particular, the mean-field constraint
\[
\langle Q_{+}\rangle \equiv \langle \hbox{s-wave pair creation operator}\rangle=0
\]
hard-wires the strong suppression of local $s$-wave pairing into the
formalism.  In the language of superconductivity, the constraint $Q_{\pm} = 0$ plays
the role of an infinite Coulomb pseudopotential, or renormalized electron-electron interaction, $\mu^{*}$. It is the satisfaction of this constraint that drives anisotropic composite pairing.

\section{The two channel Anderson lattice model} \label{Sec.2ch}

We turn now to the two channel lattice Anderson model, where the two channels
involve charge fluctuations to the ``empty'' and ``doubly'' occupied states; here we use the language of Ce$^{3+}$ whose ground state configuration is $f^1$, but this model captures any system with valence fluctuations from $f^{n-1}\rightleftharpoons
f^n \rightleftharpoons f^{n+1}$, where $n$ is odd.  For $n=1$, the empty state, $|0\rangle$ is
trivially a singlet, and we choose the doubly occupied state to be a
singlet formed from electrons in two orthogonal channels,
\begin{eqnarray}
|2\rangle = \tilde{\alpha} f\dg_{\Gamma_1,\alpha} f\dg_{\Gamma_2,  -\alpha}|0\rangle,
\end{eqnarray}  
where $|\Gamma_1: \pm\rangle \equiv |\pm\rangle$ is the ground state crystal field
doublet, and $|\Gamma_2: \pm\rangle$ is an excited crystal field
doublet.  In the finite-$U$ Anderson model, the symmetries of the
electron addition and removal processes are identical, but here
Hund's rules force the second electron to be
placed in a channel orthogonal to the first.

The local atomic Hamiltonian at site $j$ can be expressed in terms of Hubbard operators,
\begin{eqnarray}
\label{atomic}
H_a(j) & =& E_0 X_{00}(j) + E_2 X_{22}(j) + \sum_\sigma \epsilon_f X_{\sigma \sigma}(j)\cr
& + & \sum_\sigma \Delta_{CF} X_{\Gamma_2 \sigma;\Gamma_2 \sigma},
\end{eqnarray}
where the $X$'s are the Hubbard operators projecting into the empty,
$X_{00} = |0\rangle \langle 0 |$ and  doubly occupied, $X_{22} =
|2\rangle \langle 2|$ states, as well as the projectors into the 
ground state, $X_{\sigma \sigma} = |\Gamma_1 \sigma\rangle \langle \Gamma_1\sigma| \equiv |\sigma\rangle\langle \sigma|$, and excited, $X_{\Gamma_2 \sigma;\Gamma_2\sigma} = |\Gamma_2 \sigma \rangle \langle \Gamma_2 \sigma|$ crystal field doublets.   If we measure the energies from the $f$-electron level, $E_0 = -\epsilon_f$ and $E_2 = U_{12} + \epsilon_f$ are both positive, where $U_{12}$ is the Hubbard $U$ for the doubly occupied state with one electron in $\Gamma_1$ and the other in $\Gamma_2$.   We take $U_{11}, U_{22} \rarrow \infty$.  The $f$-electrons hybridize with a bath of conduction electrons, 
$H_c = \sum_{\bk,\sigma} \epsilon_\bk c\dg_{\bk\sigma} c_{\bk\sigma}$
in two different channels,
\begin{equation} 
H_V(j)\!  = \!\sum_{j} V_{1} \psi\dg_{1j\sigma} X_{0\sigma}(j)
+ V_{2}\tilde{\sigma}\psi\dg_{2j \sigma} 
X_{-\sigma 2}(j) + \text{h.c.}
\label{N2mix}
\end{equation} 
The angular momentum dependence is hidden inside the conduction Wannier states,
\begin{eqnarray}
\psi_{\Gamma j\sigma} = \sum_{\bk} \e^{-i \bk \cdot \bR_j}
\left[\Phi_{\Gamma \bk}\right]_{\sigma \sigma'} c_{\bk \sigma'},\qquad  (\Gamma=1,2)
\end{eqnarray}
where the crystal field form-factors $\left[\Phi_{\Gamma \bk}\right]_{\sigma \sigma'} = \langle k\Gamma \sigma'|\bk \sigma\rangle$ are proportional to unitary matrices.  
$X_{0\sigma} = |0\rangle \langle \Gamma_1 \sigma |$ and $X_{2\sigma} = |2\rangle \langle \Gamma_1 \sigma |$ are the Hubbard operators between the singly occupied ground state and empty or doubly occupied states, respectively.


The symplectic-$N$ Hubbard operators describing charge fluctuations in
the two channels are a simple generalization of those used for a
single channel. For a single site, these are given by 
\begin{eqnarray}\label{l}
X_{0\alpha} = b_0\dg f_\alpha + a_0\dg \tilde{\alpha} f_{-\alpha}\dg, \; X_{00} = b_0\dg b_0 + a_0\dg a_0,\cr
X_{2\alpha} = b_2\dg f_\alpha - a_2\dg \tilde{\alpha} f_{-\alpha}\dg, \; X_{22} = b_2\dg b_2 + a_2\dg a_2,
\end{eqnarray}
which can again be written more compactly by using a Nambu notation, 
\begin{eqnarray}
X_{0\alpha} & = & B_0\dg \tilde{f}_\alpha,\qquad 
X_{2\alpha}  =  B_2\dg \tilde{f}_\alpha\cr
 X_{00} &=& B_0\dg B_0, \qquad
 X_{22} = B_2\dg B_2,
\end{eqnarray}
where
\begin{equation}\label{}
B_{0} = \left(\begin{array}{c}
b_{0}\cr
a_{0}
\end{array}
\right),
\qquad 
B_{2} = \left(\begin{array}{c}
b_{2}\cr
-a_{2}
\end{array}
\right).
\end{equation}

We are free to choose the sign of $a_2$, and in the above we have
chosen the negative sign to preserve continuity with the results from
the two channel Kondo model\cite{flint08}.  The doubly occupied state is then,
\begin{eqnarray}
|2\rangle = \left(b_2\dg - a_2\dg\Psi\dg\right)|\Omega\rangle,
\end{eqnarray}
where $\Psi \dg = \sum_{\alpha >0}f\dg_{\alpha }f\dg_{-\alpha }$ is
the pair creation operator, as before. 
\fgs{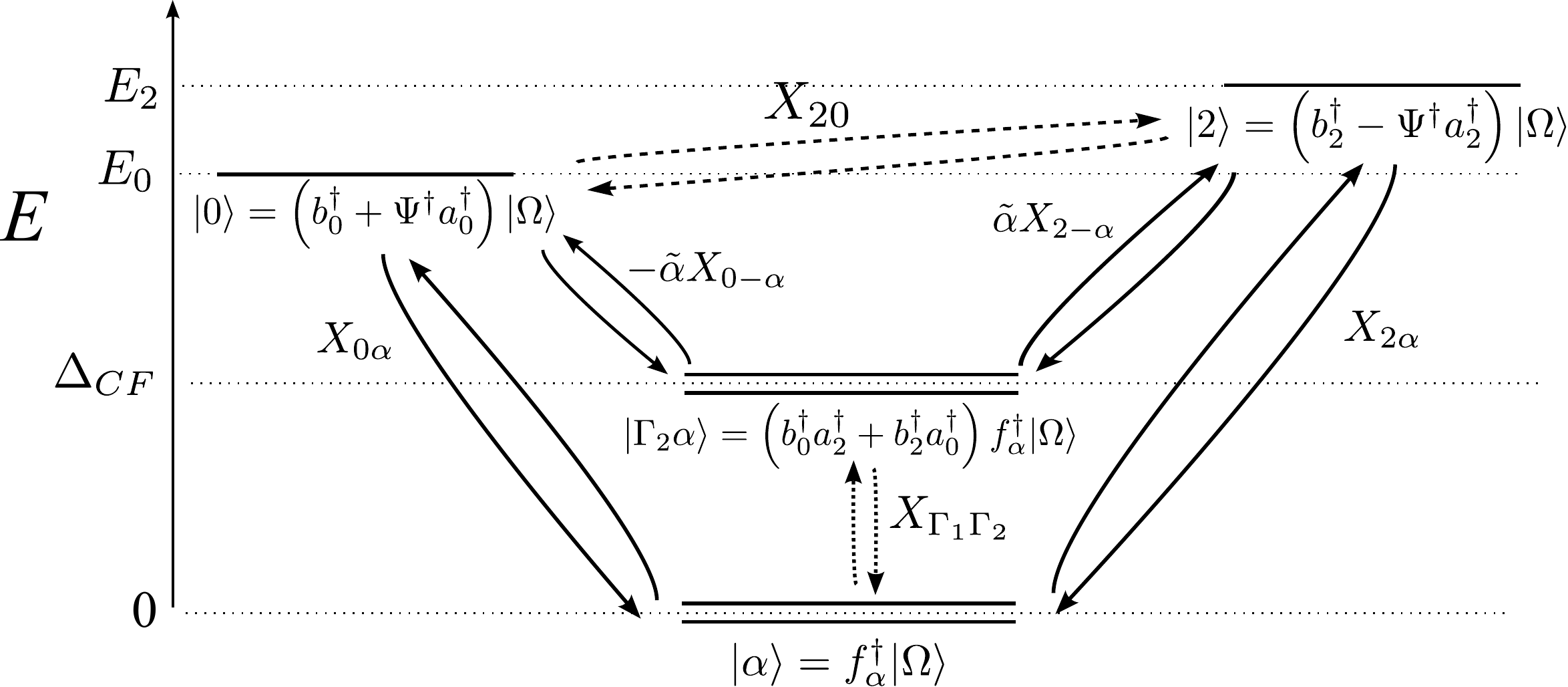}{space}{The six physical states in the 
Hilbert space of the $N = 2$ two channel Anderson model, representing
the ground state and excited crystal field doublets, and the empty and
doubly occupied singlets.  Arrows show the operators that move between
states in the Hilbert space.}

In the two channel model case the constraint becomes,
\begin{eqnarray}
\vec{Q} = B\dg_0 \vec{\tau} B_0 +B\dg_2 \vec{\tau} B_2 + \tilde{f}_\alpha\dg \vec{\tau} \tilde{f}_\alpha = 0.
\end{eqnarray}
The intersection of the two Hubbard 
algebras gives rise to an extra doublet,
\begin{eqnarray}
|\Gamma_2 \alpha\rangle = \left(b\dg_0 a_2\dg + b\dg_2 a_0\dg\right)f\dg_\alpha |\Omega\rangle,
\end{eqnarray}
which we interpret as the excited crystal field doublet because it is
reached by destroying a $\Gamma_1$ electron from the doubly occupied
state, leaving behind the $\Gamma_2$ electron.  However, this extra
doublet is not killed by $X_{00}$ or $X_{22}$, which are no longer the
projectors within this larger space.  The full Hamiltonian term for this
state must therefore be quartic in the bosons, and subtract off the
contribution from $X_{00}$, $X_{22}$ and $X_{\sigma \sigma}$,
\begin{equation}
\label{DeltaCF}
\left[\Delta_{\text{CF}}\! -\!(E_0\! +E_2\!+\epsilon_f)\right]\! \left(\! b_0\dg a_2\dg + a_0\dg b_2\dg\right)\!\left( b_0 a_2\! + a_0 b_2\right),
\end{equation}
where $\Delta_\text{CF}$ is the crystal field effect splitting.
We neglect this term for much of our analysis, as it turns out to be
irrelevant for the superconducting states of interest.

In addition to an extra state, there are two additional operators,
\begin{eqnarray}  
X_{20} & = & \{X_{2\alpha} ,X_{\alpha 0} \} = B_2\dg B_0 = b\dg_2 b_0
-a\dg_2 a_0\cr 
X_{\Gamma_1 \alpha; \Gamma_2 \beta} &=&
-\{X_{0\alpha},X_{2\beta}\} = B_0\dg(-i
\tau_2)\left[B_2\dg\right]^T\hskip -3mm \tilde{\alpha}\delta_{\alpha,-\beta}\cr
& = & \left(b\dg_0
a\dg_2 - b\dg_2 a\dg_0\right)\tilde{\alpha} \delta_{\alpha,-\beta}.
\label{operators}
\end{eqnarray} 
The operator $X_{\Gamma_1 \alpha; \Gamma_2 \beta}$ mixes the ground
state and excited crystal field doublets, leading to a composite
density wave state, $\langle c\dg_1 f f\dg c_2\rangle$.  As this state
mixes two irreducible representations of the point group, it
necessarily breaks the crystal symmetry and this phase can also be
called a composite nematic.  This mixing resembles the ordered state
proposed for URu$_2$Si$_2$[\onlinecite{haule09}], which mixes ground state and
excited singlets.

The operator $X_{20} = |2\rangle \langle 0|$ mixes the empty and
doubly occupied states, and can be thought of as a pair creation
operator.  
As we show in the next section, this operator acquires an 
expectation value when composite pairing develops.
The development of an intra-atomic order
parameter $\langle X_{02}\rangle $ is reminiscent of 
pairing in a negative (attractive)-$U$ atom\cite{anderson75}, but
unlike a negative-$U$ atom, the empty and doubly occupied sites are
{\sl excited states} of the atomic Hamiltonian,  and they are only partially
occupied as a result of valence fluctuations. Furthermore, the
paired state is an anisotropic singlet with nodes.  In the case of the Ce 115 materials, this product has $d$-wave symmetry~\cite{flint08}.  

Again, we should consider the question of supressing s-wave superconductivity.  Here,
\[
Q_+ = \sum_{\alpha>0} f\dg_\alpha f\dg_{-\alpha} + b\dg_0 a_0 + b\dg_2 a_2.
\]
We know from the one channel model that $a_0$ can be removed by a gauge transformation, eliminating the middle term, but the $b\dg_2 a_2$ term cannot be uniformly eliminated.  However, it will only be nonzero in a state where both $\langle X_{20}\rangle$ and $\langle X_{\Gamma_1 \alpha; \Gamma_2 \beta}\rangle$ are nonzero, implying coexisting superconducting and composite nematic order.  Under normal circumstances, these two phases repel one another, and s-wave superconductivity is wholly suppressed.

\section{The large $N$ two channel Anderson lattice model} \label{Sec.MF}

We are now able to write the symplectic-$N$ two channel Anderson
lattice model
in terms of the symplectic Hubbard operators, 
\begin{eqnarray}  H & = & \sum_\bk
\epsilon_k c\dg_{\bk \alpha}c_{\bk \alpha} + \sum_j E_0 B_0\dg(j)
B_0(j) + E_2 B_2\dg(j) B_2(j) \cr && + \frac{V_{1}
}{\sqrt{N/2}}\sum_{j}\psi\dg_{1 j \alpha} B_0\dg(j)\tilde{f}_{j\alpha}
+ \mathrm{h.c.} \cr
&&+ \frac{V_{2}}{\sqrt{N/2}}\sum_{j} \psi\dg_{2j \alpha}
\tilde{\alpha}\tilde{f}_{j-\alpha}\dg B_2(j)+ \mathrm{h.c.}  
\end{eqnarray}  
In order to keep the Hamiltonian extensive in $N$, we have rescaled the
hybridization terms by $(N/2)^{-1/2}$, so that they recover the
correct $N = 2$ form, (\ref{N2mix}).  This rescaling implicitly assumes
that the slave bosons fields $B_l\;(l = 0,2)$ will acquire a magntitude of order $O (\sqrt{N})$. 
Since there are only two flavors of bosons, for the bosons to play a role in the large $N $ limit, they must condense. 

Following the construction that led to the path integral description of the single-channel Anderson model (\ref{action1}) above, one can write down the Euclidean action that corresponds to this Hamiltonian:
\begin{widetext}
\begin{eqnarray}
S_2[\tPsi,\tf,\mathcal{A}_0,\mathcal{A}_2] & = & \int\limits_0^\beta \ud\tau
\sum_{\kk,{\alpha>0}}
\tPsi_{\kk{\alpha}}\dg (\pd_\tau+\eps_k
\tau_3)\tPsi_{\kk\alpha}
\label{action2} 
\cr
&&+ \sum_{j}\left\{ 
\frac{1}{2}\tr\left[ 
 \tau_3 \mathcal{A}\dg_{0j}(\pd_\tau + E_0 +
\vec{\lambda}_{j}\cdot\vec{\tau})\mathcal{A}_{0j}\right]
 + \frac{1}{2}\tr 
\left[ \tau_3
\mathcal{A}\dg_{2j}(\pd_\tau + E_2 +
\vec{\lambda}_{j}\cdot\vec{\tau})\mathcal{A}_{2j}\right]\right\}
\cr
&&+\sum_{\alpha>0}
\left\{
\tf\dg_{j{\alpha}}(\pd_\tau +
\vec{\lambda
}_{j}
\cdot\vec{\tau})\tf_{j{\alpha}} 
+  \left  ( 
\frac{V_1}{\sqrt{N/2}}
\tPsi\dg_{1j{\alpha}}\mathcal{A}_{0j}\dg\tf_{j{\alpha}}
+ \frac{V_2}{\sqrt{N/2}}
\tPsi\dg_{2j{\alpha}}\mathcal{A}_{2j}\dg\tf_{j{\alpha}} +
\text{h.c.}\right) \right\},
\label{action2}
\end{eqnarray}
\end{widetext}
\noindent where 
$\tilde{\psi }\dg_{\Gamma j {\alpha} }= (\psi \dg_{\Gamma j {\alpha} },
{\rm sgn} ( {{\alpha} }) \psi_{\Gamma j - {\alpha} })$ are  Nambu spinors
for the Wannier states in each channel.
Above, we have collected the slave bosons into the $SU(2)$ matrices,
\begin{eqnarray}
\mathcal{A}\dg_{0} = \zmatrix{b_{0}\dg}{a_{0}\dg}{a_{0}}{-b_{0}}, \quad \mathcal{A}\dg_{2} = \zmatrix{a_{2}}{b_{2}}{b_{2}\dg}{-a_{2}\dg}.
\end{eqnarray}
The action (\ref{action2}) remains manifestly invariant
under the $SU(2)$ gauge symmetry $B_l \rarrow \hR_j B_l$, where
$\hR_j$ is an $SU(2)$ matrix.  The matrices $\mathcal{A}_0,
\mathcal{A}_2$ transform under this gauge symmetry as $\mathcal{A}_l
\rarrow \hR_j \mathcal{A}_l$, leaving the product
\begin{eqnarray}
\label{A2A1}
\mathcal{A}\dg_{2}\mathcal{A}_{0} = \zmatrix{b_0 a_2 + b_2 a_0}{a_0\dg a_2 -b_0\dg b_2}{b_2\dg b_0 - a_2\dg a_0}{b_2\dg a_0\dg + a_2\dg b_0\dg}
\end{eqnarray}
gauge invariant. 

The off-diagonal components of the product
$\mathcal{A}\dg_{2}\mathcal{A}_{0}$ have the physical meaning of composite pairing between conduction and $f$-electrons. Indeed, in the Kondo limit, $E_0, E_2 \gg \pi \rho V^2$, we can connect the two channel Anderson model results to those from the two channel Kondo model\cite{flint08}.  A Schrieffer-Wolff transformation maps $b_0 \rarrow V_1$, $a_0
\rarrow \Delta_1$, $b_2 \rarrow \Delta_2$ and $a_2 \rarrow V_2$, where $V_\Gamma$ and $\Delta_\Gamma$ are the particle-hole and particle-particle hybridizations, respectively.  We can now identify the off-diagonal components of
$\mathcal{A}\dg_{2}\mathcal{A}_{0} \propto V_1 \Delta_2 - V_2 \Delta_1$ with composite
pairing. 
More generally, the components of $\mathcal{A}\dg_{2}\mathcal{A}_{0}$
can be identified with the state mixing operators,
\begin{eqnarray}
\mathcal{A}\dg_{2}\mathcal{A}_{0} = \zmatrix{X_{\Gamma_1 \Gamma_2}}{-X_{02}}{X_{20}}{X_{\Gamma_2 \Gamma_1}},
\end{eqnarray}
which confirms the identification of $\langle X_{02} \rangle$ with composite pairing, and implies that the composite pair state  contains an admixture of the empty and doubly occupied states.

To write down a translationally invariant Hamiltonian, we
assume that the expectation values of the slave bosons are
uniform, which allows us to write down the Hamiltonian in momentum
space.  To simplify this step, we temporarily drop the spin-orbit dependence of
the Wannier functions, treating the form factors as 
spin-diagonal, $\psi_{\Gamma j \alpha} = \sum_\bk c_{\bk \alpha}
\phi_{\Gamma \bk}\e^{-i \bk \cdot \bR_j}$.  This allows us to absorb
the momentum dependence of the form-factors into the hybridizations by
defining $V_{\Gamma \bk} \equiv \frac{V_\Gamma}{\sqrt{N/2}} \phi_{\Gamma
\bk}$.  To obtain the $d$-wave symmetry which arises naturally from the
spin-orbit form factors~\cite{flint08}, now we must explicitly make $V_{1k} V_{2k}$
$d$-wave.  The full spin-orbit dependence can be restored in a similar
manner to the two channel Kondo model \cite{flint08}.  The
Hamiltonian is, 
\begin{eqnarray}\label{l} 
H\! & =&\!\! \sum_{\bk \alpha}\!\!\left(\epsilon_k c\dg_{\bk
\alpha}c_{\bk \alpha}\! + V_{1\bk} c\dg_{\bk \alpha} B_0\dg
\tilde{f}_{\bk \alpha}\! + V_{2\bk} c\dg_{\bk \alpha}
\tilde{\alpha}\tilde{f}_{\bk-\alpha}\dg B_2\! +\text{h.c.}\!\right) \cr && +
\mathcal{N}_s \left(E_0 B_0\dg B_0 + E_2 B_2\dg B_2\right), \label{H2}
\end{eqnarray}
where $\mathcal{N}_s$ is the number of lattice sites. 
As before, the $SU(2)$ constraint is implemented with a vector of Lagrange
multipliers, $\vec{\lambda}$,
\begin{eqnarray}
\vec{\lambda} \cdot \left(\sum_\bk \tilde{f}\dg_{\bk \alpha} \vec{\tau} \tilde{f}\dg_{\bk \alpha} + B_0\dg \vec{\tau}B_0 + B_2\dg \vec{\tau}B_2\right),
\end{eqnarray}
where we have also assumed that $\vec{\lambda}$ is translationally
invariant, enforcing the constraint on average.  This approximation
becomes exact in the large $N$ limit.  

The Hamiltonian can be re-written in the compact form with Nambu spinors, 
\begin{eqnarray} 
\label{finalH} 
\! H \!  &=& \!\sum_{\bk \alpha}\! \left(\tilde{c}\dg_{\bk\alpha}
\; \tilde{f}\dg_{\bk\alpha}\right)\!\! \zmatrix{\epsilon_\bk
\tau_3}{V_{1\bk}\mathcal{A}_{0}\dg\!+V_{2\bk}\mathcal{A}_{2}\dg}{V_{1\bk}\mathcal{A}_{0}\!+V_{2\bk}\mathcal{A}_{2}\!}{\vec{\lambda}\cdot
\vec{\tau}}\!\! \vertvec{\tilde{c}_{\bk\alpha}}{\tilde{f}_{\bk\alpha}}
\cr & +& \left[E_0 B\dg_{0} B_{0} + E_2 B\dg_{2} B_{2} +
\vec{\lambda}\cdot
\left(B\dg_{0}\vec{\tau}B_{0}+B\dg_{2}\vec{\tau}B_{2}\right)\right],
\end{eqnarray}

We note that while in the one-channel infinite-$U$ Anderson
model (\ref{action1}), the spurious superconducting $s$-wave
state was eliminated by fixing the $SU(2)$ gauge
transformation, the two-channel model (\ref{action2}) generally
possesses a composite superconducting ground state, which cannot be
eliminated by a gauge fixing procedure. However we can always define a
transformation $\hR$ such that $a_0$ vanishes, so that $B_0\dg \rarrow
(b_0^{'\dagger }, 0)$. In this new basis, the composite pairing operator
$X_{02} = B_0\dg B_2$ (\ref{operators}) takes on the simple form, $X_{02}
= b_0^{'\dagger } b'_2$.

\subsection{The mean field solution}

In order to study the effects of mixed valence on the
superconductivity, we examine the mean field solution of the composite
pair state in the symplectic-$N$ limit.  The bosons are replaced by
their expectation values, $B_l \rarrow \langle B_l \rangle
\sim O(\sqrt(N))$, and the Hamiltonian \eqref{finalH} becomes
quadratic in the fermions, which may be integrated out exactly,
leading to the effective action $S_{eff}[B_{0},B_{2}]$ in terms of the boson
fields only. The mean-field solution is then obtained as a
saddle-point of this action: 
\beq  \frac{\delta S_{eff}[B_0,
B_2]}{\delta{B_l}} = 0.  \eeq

 We use the $SU(2)$ gauge symmetry to eliminate the $a_0$ boson, and now the composite pair state is defined by the nonzero expectation value of $\langle X_{02}\rangle = \langle b_0\dg b_2\rangle$.  While the $a_2$ boson can in principle acquire an expectation value, it would lead to a uniform composite density wave solution, which is generally unstable to the composite pair solution. Therefore in what follows, we shall set $\langle a_2\rangle=0$. 
The resulting free energy can be rewritten in terms of our mean field parameters, where we replace $\langle b_l \rangle/\sqrt{N/2}$ with $b_l$ for clarity, (however keep in mind that these have lost their dynamics),
\begin{eqnarray}
F & =&  - N T\sum_{\bk\pm} \log 2 \cosh \frac{\beta \omega_{k\pm}}{2} \cr
&& + \frac{N\mathcal{N}_s}{2} \left[b_0^2 (E_0 + \lambda_3) +b_2^2 (E_2 + \lambda_3)\right].
\end{eqnarray}
$\mathcal{N}_s$ is the number of sites, and the dispersion of the
heavy electrons is given by four branches: $\omega_{k \pm}$ and $- \omega_{k \pm}$,
where $\omega_{k \pm} = \sqrt{\alpha_k \pm \Gamma_k}$, and 
\begin{eqnarray} 
\alpha_k & = & b_+^2 + \frac{1}{2}(\epsilon_{\bk}^2 + \lambda_3^2+\lambda_1^2),\;
\Gamma_\bk = \sqrt{\alpha_k^2 -\gamma_k^2}\cr \gamma_k^2 & = &
\left[\epsilon_\bk \lambda_3 -b_-^2\right]^2 + \left[2 V_{1\bk}b_0
V_{2\bk}b_2-\epsilon_{\bk} \lambda_1\right]^2.  
\end{eqnarray}  
We have also defined 
\[
b_{\pm}^2 =
V_{1\bk}^2 b_0^2 \pm V_{2\bk}^2b_2^2.
\]
In a nodal composite pair
superconductor, the $\vec{\lambda}$ constraint reduces to $\lambda_3$,
as the $\lambda_1$ constraint acts as a Coulomb pseudo-potential
\cite{morel62} eliminating $s$-wave pairing, and it is thus unnecessary when
we choose $V_{1\bk} V_{2\bk}$ such as to give nodal superconductivity.
However, if we were to treat the finite $U$ model, where $V_{1k} =
V_{2k}$, this constraint is essential to eliminate the appearance of a
false $s$-wave superconducting phase.

The mean field parameters are determined by minimizing the free energy
with respect to $b_0$, $b_2$, $\lambda_3$ and $\lambda_1$.  To
understand their implications, we first present the mean field
equations in real space, 
\begin{eqnarray}  
&&\langle b_{1j} \rangle = \frac{V_1^2}{E_0} \sum_{\alpha}
\langle f\dg_{j \alpha} \psi_{1j \alpha} \rangle \cr
&& \langle b_{2j} \rangle =  \frac{V_2^2}{E_2}  \sum_{\alpha}\langle \tilde{\alpha}f_{j -\alpha}\psi_{2j\alpha} \rangle \cr
&&  \sum_{\alpha}\langle f\dg_{j\alpha} f_{j\alpha} \rangle  =  N/2 - \langle b_{0j}\rangle^2 - \langle b_{2j}\rangle^2 \cr 
&& \sum_{\alpha}\langle \tilde{\alpha} f\dg_{j\alpha} f\dg_{j-\alpha}\rangle  =  0.  \label{MFeqns}
\end{eqnarray}  
These equations bear a strong
resemblance to the two channel Kondo equations, where $\langle b_0
\rangle$ plays the role of the hybridization $V_1$ in channel one, 
while $\langle b_2\rangle$ plays the role of the pairing field $\Delta_2$ in channel two; here the hybridizations are explicitly
identified as the magnitude of the valence fluctuations to the empty
and doubly occupied states.  The number of singly-occupied levels,
$n_s = \sum_\alpha f\dg_\alpha f_\alpha$, is no longer fixed to $N/2$ and instead decreases as
hybridization and pairing develop.

To calculate the phase diagram, we return to the momentum space
picture, and derive the three equations relevant for composite pair
superconductivity with a nodal order parameter.  The last two equations in (\ref{MFeqns}) impose the constraint $\vec Q = 0$, fixing
$n_s = N/2 - b_0^2 -b_2^2$ and annihilating any $s$-wave pairs, while the first two equations determine the
magnitude of the valence fluctuations to the empty and doubly occupied
states, respectively:
\begin{equation}
\!\!\!\sum_\pm\!\int_\bk\! \frac{\tanh \frac{\beta \omega_{\bk \pm}}{2}}{2\omega_{\bk \pm}}\! \left\{\!\! \vertvecFour{\lambda_3}{\lambda_1}{2V_{1k}^2}{2V_{2k}^2} \pm \frac{A}{\Gamma_k}\!\!\right\} = \vertvecFour{b_0^2 + b_2^2}{0}{2(E_0+\lambda)}{2(E_2+\lambda)},
\end{equation}
where 
\begin{eqnarray}
A = \vertvecFour{\lambda_3 \alpha_k + \epsilon_\bk \left[b_-^2 - \lambda_3 \epsilon_\bk\right]}
{\lambda_1 \alpha_\bk + \epsilon_\bk\left[2 V_{1\bk} V_{2\bk} b_0 b_2 - \epsilon_\bk \lambda_1\right]}
{V_{1\bk}^2\left[(\epsilon_\bk +\lambda_3)^2+\lambda_1^2\right]+2 \lambda_1 V_{1\bk} V_{2\bk} \epsilon_\bk b_2}
{V_{2\bk}^2\left[(\epsilon_\bk -\lambda_3)^2+\lambda_1^2\right]+2 \lambda_1 V_{1\bk} V_{2\bk} \epsilon_\bk b_0 }.
\end{eqnarray}
In addition to these four equations, we must also fix the total electromagnetic charge in the system by keeping the total number of conduction electrons plus physical $f$-electrons constant.  Notice that the number of physical $f$-electrons, $n_f = N/2 - b_0^2 +b_2^2$ counts the electrons in both the singly and doubly occupied states, and differs from the occupation of the spin states, $n_s = N/2 - b_0^2 - b_2^2$ by $2b_2^2$.  

These equations can be solved numerically for a simple two-dimensional model where we take an $s$-wave $V_{1k} = V$ and a $d$-wave $V_{2k} = V \left(\cos k_x - \cos k_y\right)$, and take the two-dimensional conduction electron dispersion, $\epsilon_\bk = -2 t (\cos k_x + \cos k_y) - \mu$, where $\mu$ is adjusted to fix the total charge, $n_f +n_c$.  $\lambda_1$ can be set to zero here since by construction, we are not considering $s$-wave composite pairing.  This model contains three non-trivial phases:


\begin{itemize}
\item{For $E_0 \ll E_2$, a heavy Fermi liquid develops below the temperature  $T_1^*$.  In this phase,
$b_0$ becomes nonzero and the hybridization has symmetry $\Gamma_1$ due to  screening of $f$-spins by conduction electrons in this channel.  The $f$-electron valence will be less than $N/2$, and
these results will be identical to the infinite-$U$ Anderson model
discussed in the introduction.}
\item{For $E_2 \ll E_0$, a heavy Fermi liquid develops at $T_2^*$, where $b_2$ becomes nonzero.  This Fermi liquid will have the symmetry of the excited doublet, $\Gamma_2$ and the $f$-electron valence will be larger than $N/2$.  Again, these results should be identical to the appropriate infinite-$U$ Anderson model.}
\item{Below the superconducting temperature, $T_c$, a phase with $d$-wave composite pairing develops out of the heavy Fermi liquid, where $b_0 b_2$ becomes nonzero.  $T_c \leq \{T_1^*,T_2^*\}$ is maximal where $T_c = T_1^*=T_2^*$ and here the composite pair superconductor develops directly out of the high-temperature state of free spins, bypassing the heavy Fermi liquid.  As superconductivity is driven by the Cooper channel in the heavy electron normal states, $T_c$ is always finite and the ground state will always be superconducting.  The $f$-electron valence will generally differ from $N/2$, with nonzero contributions of both $f^0$ and $f^2$.}
\end{itemize}

\fg{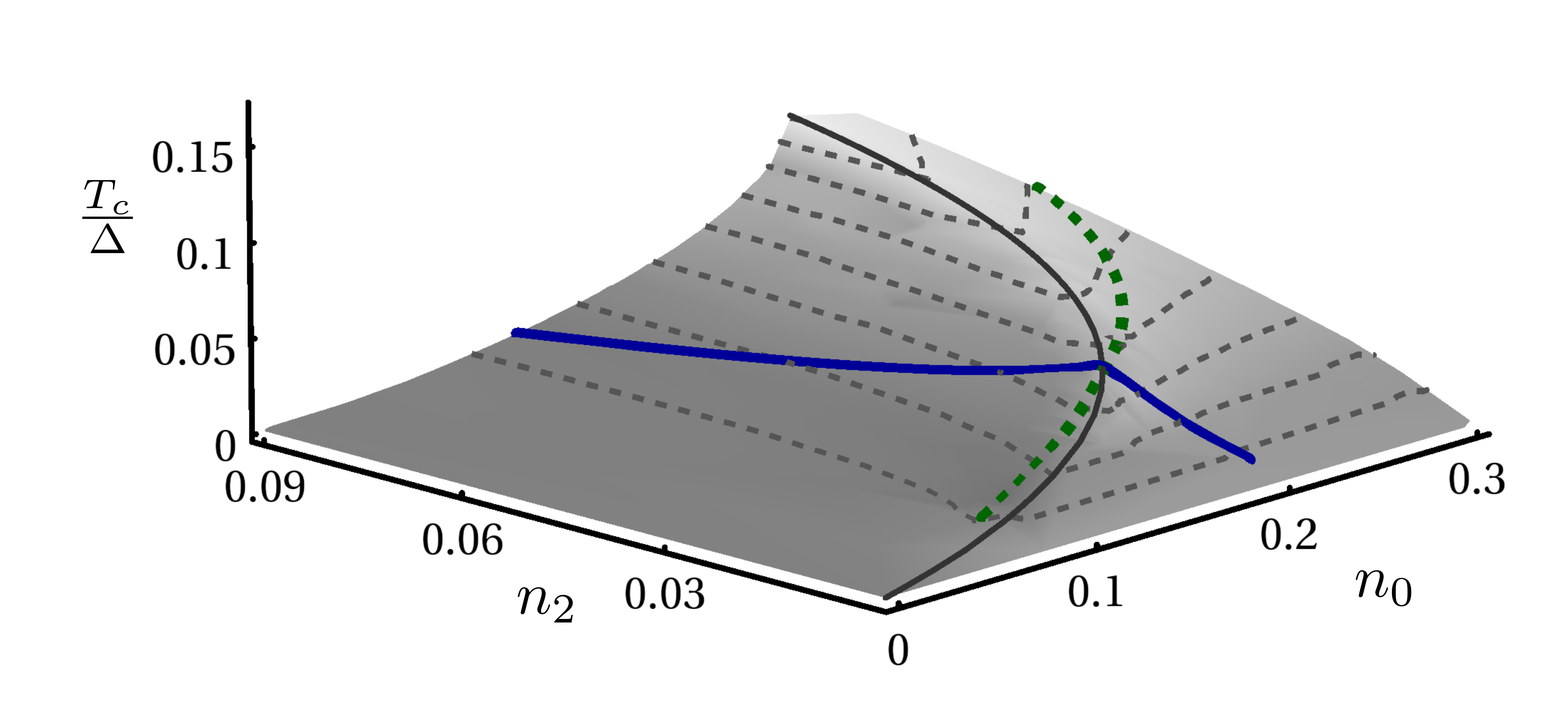}{valency}{The superconducting transition temperature is plotted against the occupancy of the empty, $n_0$ and doubly occupied, $n_2$ states, showing that $T_c$ typically increases with increasing mixed valency.  The black line indicates the maximal $T_c$, where $T_{1}^* = T_{2}^*$,  and its curve is due to the d-wave nature of the second channel.  Two different physical paths are shown: (1) The cost of doubly occupancy, $U_{12} = E_2 - E_0 - 2|\epsilon_f|$ is tuned while fixing the $f$-level, $\epsilon_f$ (blue, solid path).  Here, $E_0$ is fixed and $E_2$ increases from left to right. (2) Alternatively, the hybridization, $V$ can be increased, as by pressure or by exchanging $4f$ for $5f$ ions (green, dashed path).  
$T_c$ is scaled by $\Delta = \pi V_1^2$, and we have fixed $n_c = 1.8$ in units of $N/2$.}

\noindent
Note that in the full model, the Kondo temperatures $T_1^*$ and $T_2^*$ mark crossovers into the heavy Fermi liquids.  The appearance of phase transitions associated with condensation of $\langle b_1\rangle$ and $\langle b_2\rangle$ here is a spurious consequence of the mean-field large-$N$ treatment and is resolved with $1/N$ corrections~\cite{coleman83}.  However, the superconducting phase transition is \emph{not} spurious and will survive to finite $N$.

In Figure \ref{valency}, we plot the superconducting transition temperature versus the proportion of empty states, $n_0 = b_0^2$ and that of doubly occupied states, $n_2 = b_2^2$.  Larger $n_0$ and $n_2$ indicate a greater degree of mixed valency, which may be obtained by varying $V$, $E_0$ and $E_2$.  We show two possible paths for tuning real materials: (1) By varying $U_{12}$, $E_2=\eps_f+U_{12}$ will change, while $E_0$ remains fixed.  This path is qualitatively identical to that of the two channel Kondo model when $J_2/J_1$ is varied: there is a superconducting dome with maximal $T_c$ where $T_{1}^* = T_{2}^*$; (2) By changing the hybridization, $V$ and keeping $E_0/E_2$ fixed, $T_c$ can increase monotonically with increasingly mixed valence, as it does between CeCoIn$_5$ and PuCoGa$_5$.  A similar increase slightly away from the maximal $T_c$ can explain the non-monotonic change in $T_c$ with increasing pressure in CeCoIn$_5$[\onlinecite{sidorov02}].

\section{Charge Redistribution}\label{Sec.NQR}

As the development of composite pairing mixes the empty and doubly
occupied states, each develops a non-zero occupation, and the charge
density changes.  The link between the $f$-electron charge and the
development of the Kondo effect in the one-channel Anderson model was
explored by Gunnarsson and Schoenhammer\cite{gunnarsson83}, who showed
that the $f$-electron valence, $n_f$ decreases gradually with
temperature through the Kondo crossover, $n_f(T) = 1 - b_0^2(T)$.  The
mixing of the empty and doubly occupied states adds a new element to
this relationship, and the consideration of real, non-$s$-wave
hybridizations allows us to explore the higher angular momentum
components of the charge distribution.  The charge density can be
written, 
$\hat \rho(\bx) = \hat \psi\dg_\alpha(\bx) \hat
\psi_\alpha(\bx)$, where $\hat \psi\dg_\alpha(\bx)$ creates a physical
electron of spin $\alpha$ at $\bx$.  The  electron field $\bx$ can be
approximately decomposed as  a superposition of two orbitals 
$\Gamma_1$ and $\Gamma_2$ at nearby
lattice sites $j$,
\begin{eqnarray}
\hat \psi_\alpha(\bx) \!=\! \sum_j \left[\Phi_{1}\right]_{\alpha \beta}\!(\bx -\bR_j) \hat f_{1j\beta} + \left[\Phi_{2}\right]_{\alpha \beta}\!(\bx -\bR_j) \hat f_{2j\beta},\cr
\end{eqnarray}
where we have reintroduced the spin-orbit form factor,
$[\Phi_{\Gamma}]_{\alpha \beta}$ in order to model real materials.
The charge density of an $f$-electron located at the origin in channel
$\Gamma$ is $\rho_\Gamma(\bx) = \tr|\Phi_\Gamma(\bx)|^2 R(\bx)^2$,
where $R(\bx)$ is the radial function for the $f$-electron, and
$|\Phi_\Gamma(\bx)|^2$ is a diagonal matrix.  If we assume that the
overlap of electrons at neighboring sites is negligible, the total
charge density has three different terms,
\begin{eqnarray} 
\label{charge}
\hat \rho(\bx) & = & \sum_j \rho_{1}(\bx-\bR_j) f_{1j\beta}\dg f_{1j\beta} +\rho_{2}(\bx-\bR_j) f_{2j\beta}\dg f_{2j\beta}\cr
&& +\left[\Phi_{1}\dg \Phi_{2}\right]_{\alpha \beta}(\bx-\bR_j) f_{1j\alpha}\dg f_{2j\beta} +\mathrm{h.c.},
\end{eqnarray} 
where we have kept the spin indices of $\left[\Phi_{1}\dg \Phi_{2}\right]_{\alpha \beta}$, as it may not be diagonal in spin space.  In terms of the Hubbard operators, we can replace,  
\begin{eqnarray} 
f\dg_{1j\beta}f_{1j\beta} & = & X_{22}(j) + n_s(j) = B_{2j}\dg B_{2j} + n_s(j) \cr
f\dg_{2j \beta}f_{2j\beta} & = & X_{22}(j) + X_{\Gamma_2 \beta \Gamma_2 \beta}(j) = B_{2j}\dg B_{2j} + \gamma\dg \gamma \cr
f\dg_{1j\alpha}f_{2j\beta} & = & X_{\Gamma_1 \alpha \Gamma_2 \beta}(j),
\end{eqnarray} 
where as before, the number of singly occupied states on site $j$, $n_s(j)=N/2 - B_{0j}\dg B_{0j} - B_{2j}\dg B_{2j}$, and $\gamma\dg = B_0\dg(-i\sigma_2)[B_2\dg]^T=b_0\dg a_2\dg + b_2\dg a_0\dg $ is the operator from equation (\ref{DeltaCF}), so that $\gamma\dg f_\alpha\dg$ creates an electron in the excited single-electron crystal field state $|\Gamma_2 \alpha\rangle$, see Fig.~\ref{space}.
The third term, $X_{\Gamma_1 \alpha; \Gamma_2\beta}=|\Gamma_1,\alpha\rangle\langle\Gamma_2,\beta|$
mixes the two crystal field states, shifting charge from $\Gamma_1$ to
$\Gamma_2$. In the large-$N$ mean-field theory, this state corresponds
to non-vanishing diagonal elements of the matrix $\langle \mathcal{A}_2\dg \mathcal{A}_0
\rangle$ in Eq.~(\ref{A2A1}), $\langle X_{\Gamma_1 \alpha \Gamma_2
\beta}\rangle \sim a_0 b_2 + a_2b_0$. Since $\Gamma_1$ and $\Gamma_2$
are two different representations of the crystal point group, such
mixing implies that the crystal symmetry is spontaneously broken: in Ce-115 materials it would correspond to orthorhombic
``nematic'' density wave state, which has not been observed. We are primarily interested in
superconducting instability, and therefore ignore such a state by
setting $a_2=0$ in what follows, assuming we have already gauge fixed $a_0=0$.  Any occupation of the excited crystal field state, $\langle X_{\Gamma_2 \beta \Gamma_2 \beta}\rangle$ is also eliminated by this Ansatz.

In the heavy Fermi liquid and superconducting states, we may use the constraint to rewrite $f\dg_{1j\beta}f_{1j\beta} =  N/2 - X_{00}(j)$. The charge distribution becomes,
\begin{eqnarray}
\hat \rho(\bx)\! = \!\! \sum_j\! \rho_{1}(\bx\!-\!\bR_j\!)\!\!\left[\!\frac{N}{2}\!\! -\!\! X_{00}(j)\!\right]\!\! +\! \rho_{2}(\bx\! -\! \bR_j\!)X_{22}(j).
\end{eqnarray}
Integrating this charge density around a single site gives us the
$f$-electron valence, $n_f = 1 - b_0^2 + b_2^2$, which we plot as a
function of $T$ in Figure \ref{nf} for two different $E_2/E_0$.
The phase transition at $T^*$ is an artifact of the large $N$ limit
and becomes a crossover for any finite $N$, but the sharp kink in
$n_f$ at the superconducting transition temperature remains for all
$N$. Experimentally determining the $f$-electron valence may be possible using
core-level valence spectroscopy for Ce compounds or the M\"{o}ssbauer isomer shift for NpPd$_5$Al$_2$, where
current measurements do indicate a small, positive change in the isomer shift through $T_c$ [\onlinecite{gofryk09}].  If Np is in the $5f^3$ valence state, with dominant fluctuations between $5f^3 \rightleftharpoons 5f^4$, then the isomer shift will increase with decreasing temperature down to $T_c$, where it will begin to decrease sharply as $5f^2$ states mix in, as the black curve in Figure \ref{nf} shows.  Experimentally, the isomer shift was measured at 10K, well above $T_c$ and then below $T_c$, so the observed positive shift could just be due to the increase above $T_c$, and further measurements are necessary.
The clear observation of a sharp negative shift precisely at the superconducting transition temperature would indicate the presence of composite pairing.  

\fg{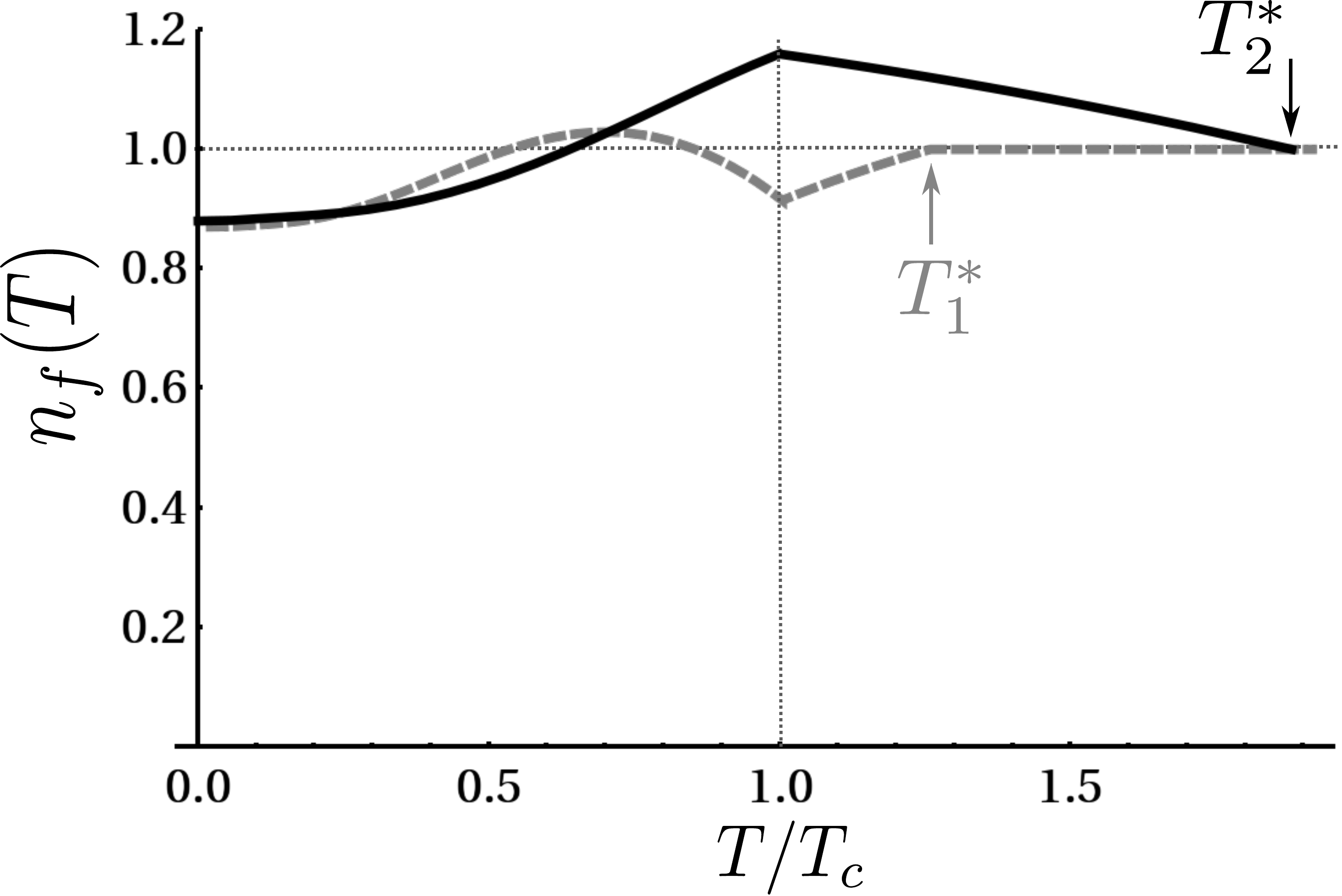}{nf}{The $f$-electron valence, $n_f$ changes as the Kondo effect and composite pair superconductivity develop.  Two examples are shown - the black curve has $T_{2}^* > T_{1}^*$, as expected for NpPd$_5$Al$_2$, so that the valence increases below $T_{2}^*$, and then sharply decreases for $T < T_c$.   For the  dashed gray curve, $T_{1}^* > T_{2}^*$, as expected for Ce$M$In$_5$ and a sharp increase in the valence is seen at $T_c$.  The sharp kinks at $T^*_1, T^*_2$ are artifacts of the mean field calculation, becoming smooth crossovers in real materials, while the superconducting kinks should be observable in real materials.}

As superconductivity develops, the occupation of the doubly occupied
state acquires an expectation value, leading to an increase in the
$\Gamma_2$ charge density. This redistribution of the $f$-electron
charge results in a quadrupole moment associated with
superconductivity.  Again, since the development of superconductivity
is a phase transition, the quadrupole moment changes sharply at $T_c$.
The quadrupole charge component has an indirect effect on the
superconducting transition temperature through its linear coupling to
strain, leading to a linear dependence of $T_c$ on the tetragonal
strain, $c/a$.  Such a linear increase of $T_c$ with $c/a$ has been
observed in both the Ce and Pu 115s, although it is conventionally
attributed to dimensionality effects\cite{bauer04}.  The quadrupole
moment of the composite condensate provides an alternate explanation,
and, in addition, leads to the development of electric field gradients
around the $f$-lattice sites, which can also be measured directly as a
shift in the nuclear quadrupole resonance frequency, $\Delta
\nu_{NQR}$ at the nuclei of the nearby atoms.

To make contact with potential experiments, we examine CeCoIn$_5$ in
more detail.  $^{115}$In atoms have a nuclear moment $I = 9/2$, which
results in a quadrupole moment, $Q = 8.3\times 10^{-29}\text{m}^2$,
making them NQR active\cite{urbano10}.  The symmetry of the ground
state doublet is $\Gamma_7^+$[\onlinecite{christianson04}], whose
angular dependence is given by,
\begin{eqnarray}
|\Phi_{\Gamma_7^+}|^2(\theta, \phi) &=& \frac{3}{16} \sin^2 \theta \left[
2 \sqrt{5} \cos 4 \phi \sin^2 \theta \sin 2 \xi \right.\\
&+& \left. 11+6 \cos 2 \xi+5 \cos 2 \theta (1+2 \cos 2 \xi)\right]\nonumber
\end{eqnarray}
where $\xi$ is a crystal-field parameter depending on the microscopic
details that can be measured with inelastic neutron scattering, and we
set $\xi \approx .25$ for CeCoIn$_5$~[\onlinecite{christianson04}].
We take the symmetry of the second channel to be $\Gamma_2 = \Gamma_6$, whose angular
dependence is,

\begin{eqnarray}
|\Phi_{\Gamma_6}|^2(\theta, \phi) = \frac{3}{32} \left[12 \cos 2 \theta +5 (3+\cos 4 \theta)\right].
\end{eqnarray}
We can now use the real charge distributions of the two orbitals to estimate the magnitude of the electric field gradient, $V_{ab} = - \partial E_a/\partial x_b$, where
\begin{eqnarray}
\rho_{\Gamma}(\bx) = |\Phi_{\Gamma}|^2(\theta, \phi) R(r)^2.
\end{eqnarray}
$R(r)$ is the $4f$ radial function,
\begin{eqnarray}
R(r) = \frac{1}{96\sqrt{35}}\left(\frac{Z^{3/2} r}{2 a_B}\right)^3 \exp\left(-\frac{Z r}{4 a_B}\right).
\end{eqnarray}
$a_B = .53$\AA $\;$ is the Bohr radius, and $Z = 6 \sqrt{10}a_B/r_{Ce}$ is adjusted so that the atomic radius is that of Ce$^{3+}$, $\langle r \rangle = r_{Ce} = 1.15$~\AA[\onlinecite{bringer83}]. 

The NQR frequency measures the electric field gradients at two different indium sites in the crystal: the in-plane, high symmetry In(1) sites, which sit in the center of a square of Ce atoms, and the lower symmetry out-of-plane In(2) sites, which are above and in-between two Ce atoms. The NQR frequency is given by\cite{white79},
\begin{eqnarray}
\nu_{NQR} = \frac{3 e V_{zz} Q}{2 h I(2I-1)} \mathrm{Hz}.
\end{eqnarray}
At the In(2) site, there is a nonzero asymmetric contribution, $\eta = |V_{xx} - V_{yy}|/V_{zz}$, which can be independently determined from experiment.
Now that we have an accurate expression for the charge distribution of the two orbitals, we may calculate the electric field gradient at $\bx$ associated with a charge $q$ in orbital $\Gamma$ at $\bR_j$:
\begin{eqnarray}
\!\!\!\!\!\!\!V_{aa}^\Gamma(\bx, \bR_j\!)\! =\!\frac{q}{4\pi \epsilon_0}\!\!\int_{\bx'}\!\!\! \rho_\Gamma(\bx'\! -\! \bR_j\!)\!\left[\! \frac{3(\!a\!-\!a')^2}{|\bx\!-\!\bx'|^5}\! -\! \frac{1}{|\bx\!-\!\bx'|^3}\!\!\right]\!.
\end{eqnarray}
Summing over the eight neighboring Ce sites is sufficient to estimate the magnitude of the NQR shift, where we use the lattice constants of CeCoIn$_5$, $a = 4.6$\AA, and $c = 7.4$\AA[\onlinecite{petrovic01a}].  The electric field gradients for a charge $e q_\Gamma$ in channel $\Gamma$ are shown in Table I.
\begin{table}
\label{EFG_tab}
\centering
\begin{tabular}{|l|c|c|}
\hline
{\bf CeCoIn$_5$} & In(1) $V_{zz}$ & In(2) $V_{zz}$  \\ \hline
$\Gamma_{7}^+$ & $-1.5 q_{7+} \cdot 10^{20}$ V/m$^2$& $+5.0 q_{7+} \cdot 10^{19}$ V/m$^2$ \\ \hline
$\Gamma_6$ &$+1.4 q_{6} \cdot 10^{20}$ V/m$^2$ & $-3.6 q_{6} \cdot 10^{19}$ V/m$^2$ \\
\hline
\end{tabular}
\caption{Estimated electronic field gradients at the two In sites in CeCoIn$_5$ due to $e q_\Gamma$ charge in the $f$-electron orbital, $\Gamma = \Gamma_{7}^+, \Gamma_6$, where $e$ is the charge of an electron.}
\end{table}

For equal channel strengths, the total charge of the $f$-ion remains unity, and the increasing occupations of the empty and doubly occupied sites cause holes to build up with symmetry $\Gamma_7^+$ and electrons with symmetry $\Gamma_6$, $q \equiv q_{6} = - q_{7+}$.  If we define $q(T)$ to be the temperature dependent occupation of the empty/doubly occupied states, in the mean field $q(T) = \langle b_0^2\rangle = \langle b_2^2\rangle$ will be proportional to $T_c - T$ just below $T_c$, and we can define,
\begin{eqnarray}
q(T) = q_0\frac{T_c - T}{T_c},
\end{eqnarray}
where $q_0$ is the ground state occupation of the empty/doubly occupied states.  In terms of $q_0$, the superconducting NQR frequency shift will be,
\begin{eqnarray}
\Delta \nu_{NQR}^1(T) & = & -7.6 q_0 \frac{(T_c - T)}{T_c} \mathrm{kHz} \cr
\Delta \nu_{NQR}^2(T) & = & +11 q_0 \frac{(T_c - T)}{T_c} \mathrm{kHz}
\end{eqnarray}
Even assuming a reasonably large $5$\% change in the single-occupancy ($q_0 = .05$) for CeCoIn$_5$  with $T_c= 2.3$K will lead to a small shift in $\nu_{NQR}^{1,2}$ with a slope of $\approx -.16, +.24$ kHz/K, beginning precisely at $T_c$.  We
could also consider the case of unequal channel strengths, where the Ce
exchanges charge with the conduction electrons, sitting in the In
$p$-orbitals, however this term will be similarly small in any clean
sample.  If this shift could be distinguished, it would be an
unambiguous signal of composite pairing.  As this shift is quite small,
M\"{o}ssbauer spectroscopy, which directly probes the $f$-ions may be a more likely technique
to observe the development of a condensate quadrupole moment.  Given that the
crystal fields for NpPd$_5$Al$_2$ are unknown, we can only make a
rough estimate of the magnitude of the quadrupolar splitting, following [\onlinecite{potzel93}].  The
$f$-ion contribution to the splitting will originate from a redistribution of charge $e q$ within the $f$-orbitals, and can be as
large as $0.2 q$~mm/s.  The lattice contribution to the quadrupole splitting, which originates from the change in $f$-valence redistributing change within the lattice, will likely be a larger effect. 

\section{Conclusions}\label{Sec.Concl}

Our two-channel Anderson model treatment has shown how composite
pairing arises as the low energy consequence of valence fluctuations in two
competing symmetry channels, which in turn manifests itself as a mixing of the
empty and doubly occupied states,
\begin{eqnarray}
\Delta_{SC} \propto \langle |0\rangle\langle2| \rangle.
\end{eqnarray} 
Composite pairing is primarily a local phenomena, where the pairing
occurs within a single unit cell.  The mixing is
reminiscent of an \emph{intra-atomic} antiferromagnetism, involving
d-wave singlet formation between the $|\Gamma_1\rangle$ and $|\Gamma_2\rangle$ states. It is 
the atomic physics of
the $f$-ions, tuned by their local chemical environment that drives
the superconductivity.  Such chemically driven d-wave pairing is a
fascinating direction for exploring higher temperature superconductors
in even more mixed valent $3d$ materials, as the strength of
the composite pairing increases monotonically with increasing valence
fluctuations, which accounts for the difference in transition temperatures
between the cerium and the actinide 115 superconductors, and for the effects of pressure.

The local nature of composite pairing should make it less sensitive to disorder than more
conventional anisotropic pairing mechanisms.  Such insensitivity is observed in the doped Ce-115 materials,
where superconductivity survives up to approximately $25\%$ substitution on the Ce site \cite{paglione07}, while neither
Ce-site nor In-site disorder behave according to Abrikosov-Gorkov, suggesting instead a percolative transition\cite{bauer10}.

The redistribution of charge due to the mixing of empty and doubly
occupied states provides a promising direction to experimentally test
for composite pairing, which should appear
as a sharp redistribution of charge associated with the
superconducting transition.  Both monopole ($f$-valence) and
quadrupole (electric field gradients) charge effects should be
observable, with core-level X-ray spectroscopy and the M\"{o}ssbauer isomer shift or as a shift in the
NQR frequency at surrounding nuclei, respectively.  We predict a shift
with slope of order $\pm0.3$kHz/K in the NQR frequency of In nuclei in CeCoIn$_5$.

Deriving these results in an exact, controlled mean field theory
required the introduction of symplectic Hubbard operators, which
maintain the time-reversal properties of $SU(2)$ electrons in the
large $N$ limit.  While our results are obtained in the large-$N$ limit, it 
should also be possible to use these Hubbard operators to
develop a dynamical mean field theory treatment of the two-channel
Anderson lattice, enabling us to examine composite pairing for $N=2$.

In addition to the two-channel Anderson model, the development of
symplectic-Hubbard operators allows a controlled treatment of the
finite-$U$ Anderson model, which is potentially useful as an impurity
solver for dynamical mean field theory.  We identify the finite-$U$
model as a special case of our two channel model when the electron and
hole fluctuations occur in the same symmetry channel, $\Gamma_1 =
\Gamma_2$.  At first sight, this model appears to give s-wave superconductivity,
however, the $SU(2)$ constraint, $\lambda_1$ forbids any on-site pairing and completely
kills the superconductivity, leaving only the simple Fermi liquid
solution.  
We expect $1/N$ corrections
to this mean field limit will differ from the $SU(N)$ approach, and an interesting future direction is to use the Gaussian fluctuations to examine the charge fluctuation side peaks.

Acknowledgements.  We should like to acknowledge discussions with
Cigdem  Capan, Maxim Dzero, Zachary Fisk, Pascoal Pagliuso and
Ricardo Urbano. 
This work was supported by the National Science Foundation, Division
of Materials Research grant DMR 0907179.

\end{document}